\documentclass[lettersize,journal,twocolumn
]{IEEEtran}
\usepackage{hyperref}       
\usepackage{url}            
\usepackage{booktabs}       
\usepackage{amsmath,amssymb,amsfonts}       
\usepackage{nicefrac}       
\usepackage{microtype}      
\usepackage{graphicx}
\usepackage{tikz}
\usetikzlibrary{shapes,positioning,plotmarks}

\def\showpgfcircle{\tikz[baseline=-0.9ex]\node[black,mark size=0.7ex]{\pgfuseplotmark{*}};}

\def\showpgfstar{\tikz[baseline=-0.9ex]\node[star,star point ratio=2.25,minimum size=6pt,inner sep=0pt,draw=black,solid,fill=black]{};}

\usepackage{acro}
\DeclareAcronym{ha}{
	short=HA,
	long= hearing aid, 
	}
 \DeclareAcronym{rm}{
	short=RM,
	long= remote microphone, 
	}
\DeclareAcronym{had}{
	short=HAD,
	long= hearing assistive device,
	}
\DeclareAcronym{snr}{
	short=SNR,
	long= signal-to-noise ratio, 
	}
\DeclareAcronym{segsnr}{
	short=segSNR,
	long= segmental signal-to-noise ratio, 
	}
\DeclareAcronym{stoi}{
	short=STOI,
	long= short-term objective intelligibility, 
	}
\DeclareAcronym{estoi}{
	short=ESTOI,
	long= extended short-term objective intelligibility, 
	}
\DeclareAcronym{mwf}{
	short=MWF,
	long= multi-channel Wiener filter, 
	}
\DeclareAcronym{wf}{
	short=WF,
	long= Wiener filter, 
	}
\DeclareAcronym{lpf}{
	short=LPF,
	long= linear prediction filter, 
	}
\DeclareAcronym{mlpf}{
	short=MLPF,
	long= magnitude linear prediction filter,
	}
\DeclareAcronym{mvdr}{
	short=MVDR,
	long= minimum variance distortion-less response,
	}
\DeclareAcronym{mpdr}{
	short=MPDR,
	long= minimum power distortion-less response,
	}
\DeclareAcronym{emwf}{
	short=EMWF,
	long= extended multi-channel Wiener filter,
	}
\DeclareAcronym{mfwf}{
	short=MFWF,
	long= multi-frame Wiener filter,
	}
\DeclareAcronym{sdmwf}{
	short=SD-MWF,
	long= speech distortion weighted multi-channel Wiener filter,
	}
\DeclareAcronym{mmse}{
	short=MMSE,
	long= minimum mean-square error,
	}
\DeclareAcronym{mse}{
	short=MSE,
	long= mean-square error,
	}
\DeclareAcronym{nmse}{
	short=nMSE,
	long= normalised mean-square error,
	}
\DeclareAcronym{lms}{
	short=LMS,
	long= least-mean-square,
	}
\DeclareAcronym{sbb}{
	short=SD-SBB,
	long= selective binaural beam-forming,
	}
\DeclareAcronym{tf}{
	short=TF,
	long= time-frequency,
	}
\DeclareAcronym{lcmv}{
	short=LCMV,
	long= linearly constrained minimum variance,
	}
\DeclareAcronym{gsc}{
	short=GSC,
	long= generalised side-lobe canceller,
	}
\DeclareAcronym{stft}{
	short=STFT,
	long= short-time Fourier transform,
	}
\DeclareAcronym{fft}{
	short=FFT,
	long= fast Fourier transform,
	}
\DeclareAcronym{atf}{
	short=ATF,
	long= acoustic transfer function,
	}
\DeclareAcronym{ratf}{
	short=RATF,
	long= relative acoustic transfer function,
	}
\DeclareAcronym{bte}{
	short=BTE,
	long= behind the ear,
	}
\DeclareAcronym{hahrir}{
	short=HAHRIR,
	long= hearing aid head related impulse response,
	}
\DeclareAcronym{rir}{
	short=RIR,
	long= room impulse response,
	}
\DeclareAcronym{wgn}{
	short=WGN,
	long= white Gaussian noise,
	}
\DeclareAcronym{hrtf}{
	short=HRTF,
	long= head related transfer function,
	}
\DeclareAcronym{hoh}{
	short=HoH,
	long= hard of hearing,
	}
\DeclareAcronym{cpsdm}{
	short=CPSDM,
	long= cross power spectral density matrix,
	}
\DeclareAcronym{cpsdms}{
	short=CPSDMs,
	long= cross power spectral density matrices,
	}
\DeclareAcronym{psd}{
	short=PSD,
	long= power spectral density,
	}
\DeclareAcronym{fir}{
	short=FIR,
	long= finite impulse response,
	}
\DeclareAcronym{wasn}{
	short=WASN,
	long= wireless acoustic sensor network,
	}
\DeclareAcronym{nfmi}{
	short=NFMI,
	long= near-field magnetic induction,
	}
\DeclareAcronym{vad}{
	short=VAD,
	long= voice activity detector,
	}
\DeclareAcronym{ssn}{
	short=SSN,
	long= speech shaped noise,
	}
\DeclareAcronym{tdoa}{
	short=TDOA,
	long= time difference of arrival,
	}
\DeclareAcronym{ifc}{
	short=IFC,
	long= inter-frame correlation,
	}
\DeclareAcronym{sos}{
	short=SOS,
	long= second-order statistics,
	}
\DeclareAcronym{bes}{
	short=BES,
	long= binary estimator selection,
	}
\DeclareAcronym{besc}{
	short=$\text{BES}_\text{C}$,
	long=  binary estimator selection for complex STFT coefficients,
	}
\DeclareAcronym{besm}{
	short=$\text{BES}_\text{M}$,
	long=   binary estimator selection for magnitude STFT coefficients,
	}
\DeclareAcronym{mstsa}{
	short=MSTSA,
	long=  multi-channel short-time spectral amplitude,
	}
\DeclareAcronym{stsa}{
	short=STSA,
	long= short-time spectral amplitude,
	}
\DeclareAcronym{mushra}{
	short=MUSHRA,
	long= multiple stimuli with hidden reference and anchor,
	}
\DeclareAcronym{srt}{
	short=SRT,
	long= speech reception threshold,
	}
\DeclareAcronym{asr}{
	short=ASR,
	long= automatic speech recognition,
	}
\DeclareAcronym{ver}{
	short=VER,
	long= vent external response,
	}
\DeclareAcronym{mccc}{
	short=MCCC,
	long= multi-channel cross correlation,
	}
\DeclareAcronym{ncc}{
	short=NCC,
	long= normalized cross correlation,
	}
\DeclareAcronym{mog}{
	short=MOG,
	long= minimum overlap-gap,
	}
\DeclareAcronym{eeg}{
	short=EEG,
	long= electroencephalogram,
	}
\DeclareAcronym{eog}{
	short=EOG,
	long= electroocoulogram,
	}
\DeclareAcronym{drr}{
	short=DRR,
	long= direct-to-reverberant ratio,
	}
\DeclareAcronym{hats}{
	short=HATS,
	long= head-and-torso simulator,
	}
\DeclareAcronym{cs}{
	short=CS,
	long= channel selection,
	}
\begin{document}

\title{Head-steered channel selection method for hearing aid applications using remote microphones}

\author{Vasudha Sathyapriyan \qquad Michael S. Pedersen \qquad Mike Brookes \\ Jan \O{}stergaard \qquad Patrick A. Naylor \qquad Jesper Jensen
\thanks{This project has received funding from the European Union’s Horizon 2020
research and innovation programme under the Marie Skłodowska-Curie grant
agreement No. 956369.}
\thanks{Vasudha Sathyapriyan, Michael S. Pedersen and Jesper Jensen are with Demant A/S, 2765 Smørum, Denmark (e-mail: {vasy, micp, jesj}@demant.com). Vasudha Sathyapriyan and Jesper Jensen are also with the Department of Electronic Systems, Aalborg University, 9220 Aalborg, Denmark}
\thanks{Mike Brookes and Patrick A. Naylor are with the Department of Electrical and Electronic Engineering, Imperial College London, London SW7 2AZ, United Kingdom (e-mail: {mike.brookes, p.naylor}@imperial.ac.uk).}
\thanks{Jan Østergaard is with the Department of Electronic Systems, Aalborg University, 9220 Aalborg, Denmark (e-mail: jo@es.aau.dk).}}

\maketitle

\begin{abstract}
We propose a channel selection method for hearing aid applications using remote microphones, in the presence of multiple competing talkers. The proposed channel selection method uses the hearing aid user's head-steering direction to identify the remote channel originating from the frontal direction of the hearing aid user, which captures the target talker signal. We pose the channel selection task as a multiple hypothesis testing problem, and derive a maximum likelihood solution. Under realistic, simplifying assumptions, the solution selects the remote channel which has the highest weighted squared absolute correlation coefficient with the output of the head-steered hearing aid beamformer. We analyze the performance of the proposed channel selection method using close-talking remote microphones and table microphone arrays. Through simulations using realistic acoustic scenes, we show that the proposed channel selection method consistently outperforms existing methods in accurately finding the remote channel that captures the target talker signal, in the presence of multiple competing talkers, without the use of any additional sensors.
\end{abstract}


\section{Introduction}\label{sec:Introduction}
Microphone arrays are used in various speech and audio signal processing applications, due to their ability to exploit the spatial diversity of the acoustic scene to selectively enhance the signals appearing from a chosen direction \cite{brandstein_microphone_2001}. They are typically used in applications such as \acp{ha} \cite{doclo2015multichannel}, \acp{wasn} \cite{bertrand2011applications}, \ac{asr} \cite{wolfel2009distant}, and teleconferencing systems \cite{brandstein_microphone_2001}. 

The performance of microphone arrays depends not only on the number of microphones but also on the \ac{snr} and the \ac{drr} of the target signal captured by them \cite{brandstein_microphone_2001}. In particular, when the microphones are positioned far from the target talker, they tend to capture low \ac{snr} and low \ac{drr} signals, limiting their noise reduction performance. In applications such as \acp{ha}, where microphone placement is constrained by the device size, only the local acoustic field around the user's head is sampled, thus limiting noise reduction compared to using a \ac{rm} positioned closer to the target source \cite{thibodeau2014comparison,boothroyd2004hearing,thibodeau2020benefits,de2016speech}. To address this, it has been proposed to employ multiple \acp{rm} in \acp{wasn}, where microphones are distributed across the acoustic scene, covering a larger physical area \cite{bertrand2011applications,kates2019integrating,gossling2020binaural}. This approach can capture the target signal at a higher \ac{snr} and \ac{drr}, even in complex, dynamic acoustic environments. 

However, using multiple \acp{rm} comes with the demand for more power, bandwidth, and computational resources, the availability of which can be limited, especially in small, battery-powered devices such as \acp{ha}. In \cite{kumatani2011channel, bertrand_robust_2009,bertrand2010efficient}, the authors show that in \acp{wasn}, using a particular subset of \ac{rm} channels can achieve essentially the same performance as using all available \ac{rm} channels in the room. In particular, using the sensor with the microphone channel located closest to the target talker significantly enhances the target talker signal. Methods using utility functions, which measure the cost of adding or removing a remote sensor to the signal estimation task, have been used for remote \ac{cs} \cite{bertrand2010efficient,bertrand2012efficient,szurley2012efficient,szurley2012energy,szurley2014greedy,zhang2017microphone,zhang2021sensor,gunther_network_aware,gunther2021online,gunther2023microphone}. In \cite{bertrand2010efficient,bertrand2012efficient,szurley2012efficient,szurley2012energy,szurley2014greedy}, the authors used \ac{mmse} utility function for the signal estimation task. However, these methods assume knowledge of inter-channel \ac{sos} related to the target talker signal, or estimate them under the assumption of a single-talker environment, i.e., when the acoustic scene consists of only the target talker in noise, but with no competing talkers. Furthermore, in \cite{zhang2017microphone,zhang2021sensor,gunther_network_aware,gunther2021online,gunther2023microphone}, the authors proposed network-aware optimal microphone selection methods. Although the work in \cite{zhang2017microphone,zhang2021sensor} focused on minimizing communication cost while limiting output noise power, the authors in \cite{gunther_network_aware,gunther2021online,gunther2023microphone} proposed network-aware microphone selection methods that are not limited to a specific application. However, these methods either assume that the target talker signal is the only coherent signal captured by all the remote channels, or used the knowledge of its location to characterize it as the dominant source in a multi-talker environment, i.e., when the acoustic scene consists of competing talkers in addition to the target talker.

In the context of automatic speech recognition, various remote \ac{cs} strategies have been proposed based on choosing the \ac{rm} channel with the highest \ac{snr}\cite{wolfel2006multi}, the \ac{rm} channel with the highest \ac{drr} \cite{wolf2014channel}, the \ac{rm} channels with the maximum cross-correlation with each other \cite{kumatani2011channel}, or by clustering \ac{rm} channels according to their proximity to the target talker \cite{himawan2010clustering}. However, \ac{snr}-based methods are sensitive to the quality of the \acp{vad} in noisy environments, and a good estimation of the \ac{rir} can be challenging in highly dynamic environments for the \ac{drr}-based method. Furthermore, the \ac{mccc} method \cite{kumatani2011channel} requires prior knowledge of the \ac{rm} positions and, more importantly, relies on the limiting assumption of a single-talker environment, i.e., no competing talkers are present.

The requirement of a single-talker environment or knowing the target talker in a multi-talker environment limits the use of existing methods \cite{bertrand2010efficient,bertrand2012efficient,szurley2012efficient,szurley2012energy,szurley2014greedy,zhang2017microphone,zhang2021sensor,gunther_network_aware,gunther2021online,gunther2023microphone}, for \ac{ha} applications. In fact, \ac{ha} users are commonly surrounded by multiple talkers, e.g., in restaurants, offices, and public transport. In such situations, the remote \ac{cs} algorithm must first identify the target talker among the set of competing talkers, which is the problem addressed in this paper. 

The problem of identifying the target talker in a multi-talker environment in a \ac{ha} application is essentially ill-posed: it is very hard to identify the target talker without additional information about the target talker, e.g., physical location, signal characteristics, or the \ac{ha} user's intention. The problem has previously been addressed using the \textit{turn-taking} behavior of humans \cite{sacks1978simplest,mccowan2003modeling}, using \textit{auditory attention decoding} \cite{alickovic2019tutorial,geirnaert2021electroencephalography,mirkovic2016target}, or simply via using a manual interface. \textit{Turn-taking} based methods detect the target talker using speech gaps and overlaps in the conversation between the \ac{ha} user and multiple candidate talkers in the acoustic scene, but they require robust \acp{vad} of candidate talkers and that the user is engaged in active conversation rather than just listening \cite{harma2009conversation,hoang2022minimum}. Methods based on \textit{auditory attention decoding} use \ac{eeg} or \ac{eog} electrodes to detect the \ac{ha} users' attention or to monitor the \ac{ha} user's eye-gaze to detect the target talker. Although various method have been proposed that use \ac{eeg} to detect the \ac{ha} user's attention, the methods rely largely on the availability of wet \ac{eeg} measurements recorded using bulky scalp electrodes \cite{geirnaert2021electroencephalography}. Although methods using miniature electrodes in/ around the ear, the so-called ear-\ac{eeg} \cite{ciccarelli2019comparison, kam2019systematic}, increase their wearability factor, the development of wearable, robust \ac{eeg} based \acp{ha} still remains a challenge. Lastly, selecting the target talker via a manual interface is less practical, particularly in dynamic acoustic situations.

Motivated by the recent development of wireless microphone accessories for \acp{ha} such as, clip-on microphones, e.g., \cite{OticonConnectClip,PhonakRogerPen}, and table microphones, e.g., \cite{PhonakTable}, that are designed for \ac{ha} users in challenging situations, e.g., social gatherings, our work focuses on \ac{cs} in multi-talker situations, where the \ac{ha} users have access to such \ac{rm} accessories.

People, including \ac{ha} users, use lip-reading to improve speech intelligibility and lower \acp{srt} \cite{middelweerd1987effect,erber1969interaction,bernstein2022lipreading}. Moreover, \ac{ha} users are recommended by hearing care professionals to lip read in social situations \cite{valente2006guideline,chandrasekhar2019clinical,turton2020guidelines}. Therefore, \ac{ha} users tend to, often, steer their heads toward their conversational partners in social settings. This fact has formed the basis for the frontal target talker assumptions in multi-microphone speech enhancement methods for \ac{ha} applications \cite{doclo2015multichannel}.

In this paper, we use the head-steering direction of the \ac{ha} user to address the problem of remote \ac{cs} from a distributed network of \ac{rm} channels, in a dynamic, multi-talker environment, where in addition to the target talker, multiple competing talkers are present, without additional prior knowledge about the target talker location, signal characteristics  or using additional sensors. We pose the \ac{cs} task as a classification problem and derive an analytical solution using a multiple hypothesis test framework. We verify experimentally that the proposed method selects the remote channel that captures the target talker located along the head-steering direction, and consequently, selects the channel with the highest \ac{snr}. The results indicate that the proposed method outperforms state-of-the-art methods, in accurately identifying the remote channel. Moreover, we also demonstrate that the performance of the proposed method is robust to mismatches in the frontal \acp{ratf} and analyze the performance of the proposed methods for non-frontal target talkers. Lastly, we illustrate a practical application of the proposed method in a conference room setting, where table microphone arrays are equipped with beamformers steered towards fixed locations. We show that the proposed method can accurately select the output of a table microphone beamformer that enhances the target talker, located in the head-steering direction of the \ac{ha} user. 

To limit the scope of this work, we consider the situation where multiple \ac{rm} channels are transmitted to the \ac{ha} as the fusion center. Challenges related to the transmission of multiple wireless channels to the \ac{ha} such as, limited transmission power, bit-rate allocation, sample rate mismatch in the devices, are outside the scope of this study and we direct the reader to existing studies for more detailed discussions: \cite{gunther_network_aware,zhang2019sensor,zhang2022energy}. Furthermore, in deriving the proposed method, we assume that the \ac{ha} user steers their head in the direction of the target talker. In a real world situation, this may not constantly be the case, e.g., due to head movements of the user. However, such movements can relatively easily be devised, but is outside the scope of the paper. Besides, as we show in Section \ref{subsec:NonFrontal_RATF_analysis}, the proposed method is relatively insensitive to this assumption.

The structure of the paper is as follows. In Section \ref{sec:SignalModel} we present the signal model and notations that will be used in deriving the proposed method. In Section \ref{sec:propMethod}, we derive analytical expressions for the proposed \ac{cs} method. In Section \ref{sec:Close_RM}, we analyze the performance of the proposed method through simulations and demonstrate practical applications. Lastly, we summarize and conclude our work in Section \ref{sec:Conclusion}.

\section{Signal Model \& Notations}\label{sec:SignalModel}
\begin{figure}[htbp]
    \centering
    \includegraphics[width=\linewidth]{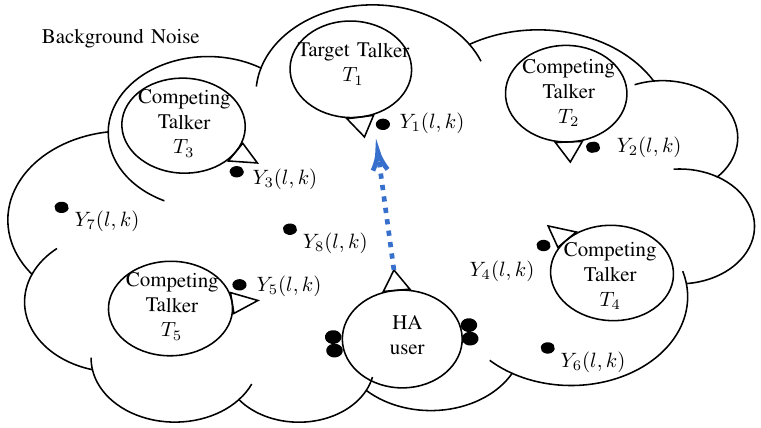}
    \caption{Example acoustic scene with a \ac{ha} user, with $M\geq 2$ microphones, and with $N+1$ candidate talkers, with $R$ \acp{rm} channels distributed around the scene. The \ac{ha} user's head is steered towards the target talker, $T_1$. All the microphones are denoted by `\protect\showpgfcircle'.}
    \label{fig:SignalModel}
\end{figure}

Consider an acoustic setup, with a \ac{ha} user wearing \acp{ha}, with ${M} \geq 2$ microphones, surrounded by $N+1$ candidate talkers, see Figure \ref{fig:SignalModel}. Let $T_1$ be the target talker, while the candidate talkers, $T_i$, $i\geq 2$, are the $N$ competing talkers. Let there be remote acoustic sensors distributed around the acoustic scene that capture and transmit audio signals to the \ac{ha} device. The remote acoustic sensors could be single \acp{rm} or microphone arrays, and the signals sent to the \ac{ha} could be unprocessed or processed signals from each sensor. For simplicity, and without loss of generality, in this section, we assume the remote sensors to be \acp{rm}. The \acp{rm} can either be close-talking \acp{rm} located near the candidate talkers, e.g., clip-on microphones, or located elsewhere in the room, see Figure \ref{fig:SignalModel}. 

The \ac{stft} coefficient vector of the noisy signal captured by the \ac{ha} microphones, $\mathrm{\mathbf{y}_{HA}}(l,k) \in \mathbb{C}^{M \times 1}$, is given by

\begin{equation}
    \begin{split}
    \mathrm{\mathbf{y}}_\mathrm{HA}(l,k) &=  \mathrm{\mathbf{a}}(\theta_1,k) S_1(l,k)\\ &+  \underbrace{\sum_{i = 2}^{N+1}\mathrm{\mathbf{a}}(\theta_i,k) S_i(l,k) +\mathrm{\mathbf{v}_{b}}(l,k)}_{\mathrm{\mathbf{v}_{HA}}(l,k)},
    \end{split}
    \label{eq:HA_model}
\end{equation}
where, $S_i(l,k) \in \mathbb{C}$ is the \ac{stft} coefficient of the speech signal from candidate talker $T_i$, $\mathbf{a}(\theta_i,k)\in \mathbb{C}^{M \times 1}$ is the head-related \ac{atf} vector from the $i^\mathrm{th}$ candidate talker to all $M$ \ac{ha} microphones, $\mathbf{v}_\mathrm{b}(l,k)\in \mathbb{C}^{M \times 1}$ is the \ac{stft} coefficient vector of the background noise component at the \ac{ha} microphones, and $\mathrm{\mathbf{v}_{HA}}(l,k)\in \mathbb{C}^{M \times 1}$ is the \ac{stft} coefficient vector of the total noise component at the \ac{ha} microphones, defined as the sum of the competing talkers and the background noise.
Let $\mathbf{d}(\theta_i,k)\in \mathbb{C}^{M \times 1}$, be the \ac{ratf} from the $i^\mathrm{th}$ candidate talker to the \ac{ha} microphones, measured w.r.t. an arbitrarily chosen \ac{ha} reference microphone, i.e.,  $\mathbf{d}(\theta_i,k) = \frac{\mathbf{a}(\theta_i,k)}{A_\mathrm{ref}(\theta_i,k)}$ \cite{gannot2001signal}, where $A_\mathrm{ref}(\theta_i,k)\in \mathbb{C}$, is the \ac{atf} from the $i^\mathrm{th}$ candidate talker to the \ac{ha} reference microphone. We can then rewrite (\ref{eq:HA_model}) as
\begin{equation}
    \begin{aligned}
    \mathrm{\mathbf{y}_{HA}}(l,k) &=  \mathrm{\mathbf{d}}(\theta_1,k) S_\mathrm{1,ref}(l,k)+  \mathrm{\mathbf{v}_{HA}}(l,k),
    \end{aligned}
    \label{eq:HA_model_RTF}
\end{equation}
where $S_\mathrm{1,ref}(l,k)\in \mathbb{C}$ is the \ac{stft} coefficient of the target speech signal at the \ac{ha} reference microphone. We model the \ac{ha} noise signal vector, $\mathbf{v}_\mathrm{HA}(l,k)$ as a zero-mean, complex Gaussian distributed random vector with a \ac{cpsdm}, $\underline{\mathbf{C}}_\mathbf{\mathrm{v}}(l,k)$, defined as
\begin{equation}
    \begin{aligned}
        \underline{\mathbf{C}}_\mathbf{\mathrm{v}}(l,k) \triangleq \mathbb{E}\left[\mathrm{\mathbf{v}_{HA}}(l,k)\mathrm{\mathbf{v}}_\mathrm{HA}^H(l,k)\right] \in\mathbb{C}^{M \times M},
    \end{aligned}
    \label{eq:HA_noisy_cov}
\end{equation}
and the \ac{ha} noisy \ac{cpsdm} is defined as
\begin{equation}
    \begin{aligned}
        \underline{\mathbf{C}}_\mathbf{\mathrm{y}}(l,k) \triangleq \mathbb{E}\left[\mathrm{\mathbf{y}_{HA}}(l,k)\mathrm{\mathbf{y}}_\mathrm{HA}^H(l,k)\right] \in\mathbb{C}^{M \times M}.
    \end{aligned}
    \label{eq:HA_noise_cov}
\end{equation}

Let there be $R$ \acp{rm} placed in the acoustic scene. The \ac{stft} coefficient of the $r^\mathrm{th}$ noisy \ac{rm} signal, $Y_r(l,k)\in \mathbb{C}$, is given by
\begin{equation}
    \begin{split}
    Y_r(l,k&)\!= \! A_{1,r}^\prime(k) S_{1}(l,k) \\ &+ \! \underbrace{ \sum_{i=2}^{N+1} \! A_{i,r}^\prime(k) S_{i}(l,k) \! + \! V_{\mathrm{b},r}(l,k)}_{V_r(l,k)}, \,  r \!=\! 1,\hdots,R,    
    \end{split}
    \label{eq:RM_model}
\end{equation}
where $A_{i,r}^\prime(k,l)\in \mathbb{C}$ is the \ac{atf} from the $i^\mathrm{th}$ candidate talker to the $r^\mathrm{th}$ \ac{rm}, $V_{\mathrm{b},r}(l,k)$ is the \ac{stft} coefficient of the background noise component at every $r^\mathrm{th}$ \ac{rm} and $V_r(l,k)$ is the \ac{stft} coefficient of the total noise component at the $r^\mathrm{th}$ \ac{rm}, where the noise is the sum of the competing talkers and the background noise. 

\section{Proposed Method}\label{sec:propMethod}
\subsection{Problem formulation and Solution}\label{subsec:ProblemFormulation}
We pose the task of \ac{cs} based on the head-steering direction of the \ac{ha} user, as a problem of classifying which $r^\mathrm{th}$ channel contains the target talker signal, and fits the signal model of the \ac{ha} microphone signal. Specifically, we formulate the following multiple hypothesis testing problem 
\begin{equation}
    \begin{split}
    \mathcal{H}_r(l,k) : \mathrm{\mathbf{y}_{HA}}(l,k) &=  \mathrm{\mathbf{d}}(0^\circ,k) A_{1,r}(k) Y_r(k,l) \\ &+ \mathrm{\mathbf{v}_{HA}}(l,k), \quad r = 1,\hdots, R,
    \end{split}
    \label{eq:Hypotheses}
\end{equation}
where $\mathcal{H}_r(l,k)$ denotes the hypothesis that the $r^\mathrm{th}$ channel contains the target talker signal, $\mathrm{\mathbf{d}}(0^\circ,k)$, is the \ac{ratf} from the frontal target talker, measured at the \ac{ha} microphones relative to the \ac{ha} reference microphone, $A_{1,r}(k) \triangleq \frac{A_\mathrm{ref}(k)}{A_{1,r}^\prime(k)}$ is the \ac{ratf} of the target talker, measured at the \ac{ha} reference microphone, relative to the $r^\mathrm{th}$ \ac{rm}. We assume all the \acp{atf} to be shorter than the frame length of the \ac{stft}. Unlike in many existing speech enhancement schemes which require short \ac{stft} frames to limit the algorithm delay \cite{gannot2001signal,doclo2015multichannel}, we can use longer frame-lengths here, as the proposed method is only used for remote \ac{cs}. 

In (\ref{eq:Hypotheses}), by using a front-fixed \ac{ratf}, $\mathrm{\mathbf{d}}(0^\circ,k)$, we model the assumption that the \ac{ha} user's head is steered towards the target talker. For the sake of clarity, let us briefly consider a constrained situation where the \acp{rm} in Figure \ref{fig:SignalModel} are close-talking \acp{rm}, such that the \acp{rm} capture only the clean signals from the candidate talkers, $T_i$, for $1\!\leq \! i \! \leq \! 5$, such that $Y_i(l,k) \!=\! A_{i,i}^\prime(k) S_i(l,k)$ in (\ref{eq:RM_model}). Then, $Y_1(l,k)$, captured by the \ac{rm} placed on the target talker would perfectly fit the model in (\ref{eq:Hypotheses}). Returning to the acoustic scene in Figure \ref{fig:SignalModel}, when the \acp{rm} may be moved away from the candidate talkers, in addition to the target talker signal, the \acp{rm} would capture the competing talker signals and background noise, as in (\ref{eq:RM_model}). This reduces the classification hypothesis in (\ref{eq:Hypotheses}), to finding the \ac{rm} closest to the target talker,  who is located in the head-steering direction of the \ac{ha} user.

Assuming equal prior probabilities of each hypothesis, $\mathcal{H}_r(l,k)$, the problem of minimising the probability of error of classification becomes a maximum likelihood problem \cite{kay1993fundamentals}. Therefore, we select the $r^\mathrm{th}$ hypothesis, $\mathcal{H}_r(l,k)$, with the highest conditional likelihood $p(\mathrm{\mathbf{y}_{HA}}(l,k)| \mathcal{H}_r)$, and the channel selected is given by
\begin{equation}
    \begin{aligned}
    \hat{r}_{\mathrm{prop}}(l,k) = \operatorname*{argmax}_{r \in \{1, \hdots, R\}} p(\mathrm{\mathbf{y}_{HA}}(l,k)| \mathcal{H}_r(l,k)).
    \end{aligned}
    \label{eq:Selection}
\end{equation}
In (\ref{eq:Hypotheses}), $\mathrm{\mathbf{d}}(0^\circ,k)$, $Y_r(l,k)$ are known, deterministic variables, $\mathrm{\mathbf{v}_{HA}}(l,k)$ is assumed to be zero-mean complex Gaussian distributed random vector with a \ac{cpsdm}, $\underline{\mathbf{C}}_\mathbf{\mathrm{v}}(l,k)\in\mathbb{C}^{M \times M}$, so that $\mathrm{\mathbf{y}_{HA}}(l,k)$ is a non-zero mean, complex Gaussian distributed random vector. The conditional likelihood function, $p(\mathrm{\mathbf{y}_{HA}}(l,k)| \mathcal{H}_r(l,k))$, parameterized by $\phi_r(l,k) = \{Y_r(l,k)$, $\mathrm{\mathbf{d}}(0^\circ,k)$, $\underline{\mathbf{C}}_\mathbf{\mathrm{v}}(l,k)$, $A_{1,r}(k)\}$, is then given by
\begin{equation}
    \begin{aligned}
    &p(\mathrm{\mathbf{y}_{HA}}(l,k)| \mathcal{H}_r(l,k),\phi_r(l,k)) =\frac{1}{\pi^M \det{[\underline{\mathbf{C}}_\mathbf{\mathrm{v}}(l,k)}]}\\
    &\quad \times \exp\Bigl(-\left[ \mathrm{\mathbf{y}_{HA}}(l,k) - \mathrm{\mathbf{d}}(0^\circ,k)\mu_r(l,k)\right]^H\\
    & \quad \quad \quad \quad \times \underline{\mathbf{C}}_\mathbf{\mathrm{v}}^{-1}(l,k)\\
    & \quad \quad \quad \quad \times \left[ \mathrm{\mathbf{y}_{HA}}(l,k) - \mathrm{\mathbf{d}}(0^\circ,k)\mu_r(l,k)\right]\Bigr),
    \end{aligned}
    \label{eq:cond_likelihood}
\end{equation}
where $\mu_r(l,k) \triangleq A_{1,r}(k) Y_r(l,k)$.

Let the \ac{ha} noisy signal data matrix of $D$ noisy \ac{ha} observations for a given frequency bin, $k$, be given by
\begin{equation}
    \begin{aligned}        \mathrm{\overline{\underline{\mathbf{y}}}_{HA}}(l,k) = \begin{bmatrix}
\mathrm{\mathbf{y}_{HA}}(j_l,k) & \hdots & \mathrm{\mathbf{y}_{HA}}(j_u,k)
\end{bmatrix},
    \end{aligned}
    \label{eq:multi_observations_HA}
\end{equation}
where $j_l \!=\! l-D+1$ and $j_u = l$, and let the noisy data matrix, for all $K$ frequency bins be
\begin{equation}
    \begin{aligned}         \mathrm{\overline{\overline{\underline{\underline{\mathbf{y}}}}}_{HA}}(l) = \begin{bmatrix}
\mathrm{\overline{\underline{\mathbf{y}}}_{HA}}(l,1) & \hdots & \mathrm{\overline{\underline{\mathbf{y}}}_{HA}}(l,K)
\end{bmatrix}.
    \end{aligned}
    \label{eq:multi_freq_HA}
\end{equation}
Assuming that the \ac{stft} observations are independent across time-frame, $l$ and frequency-bins, $k$, the joint conditional likelihood function is given by
\begin{equation}
    \begin{split}
&p(\mathrm{\overline{\overline{\underline{\underline{\mathbf{y}}}}}_{HA}}(l)| \mathcal{H}_r(l,k),\phi_r(l,k)) \\ &\quad = \prod_{k=1}^{K} \prod_{j = j_l}^{j_u} p\Bigl(\mathrm{\mathbf{y}_{HA}}(j,k)| \mathcal{H}_r(j,k),\phi_r(j,k)\Bigr).
    \end{split}
    \label{eq:joint_cond_likelihood}
\end{equation}
Using the natural logarithm on both sides of (\ref{eq:joint_cond_likelihood}), we get
\begin{equation}
    \begin{split}
&\ln\Bigl[{p(\mathrm{\overline{\overline{\underline{\underline{\mathbf{y}}}}}_{HA}}(l)| \mathcal{H}_r(l,k),\phi_r(l,k))}\Bigr] \\ 
&\quad =  \sum_{k = 1}^{K} \sum\limits_{j =j_l}^{j_u} \ln p\Bigl(\mathrm{\mathbf{y}_{HA}}(j,k)| \mathcal{H}_r(j,k),\phi_r(j,k)\Bigr)\\
&\quad = -KDM \ln{\pi} -\sum_{k = 1}^{K}\sum\limits_{j = j_l}^{j_u} \ln\Bigl(\det{[\underline{\mathbf{C}}_\mathbf{\mathrm{v}}(j,k)]}\Bigr) \\
&\quad -\sum_{k=1}^{K}\sum\limits_{j=j_l}^{j_u} \mathrm{\mathbf{y}_{HA}}^H(j,k)\underline{\mathbf{C}}_\mathbf{\mathrm{v}}^{-1}(j,k) \mathrm{\mathbf{y}_{HA}}(j,k) \\ 
&\quad + \sum_{k=1}^{K} \! \sum\limits_{j=j_l}^{j_u} \! \Bigl[\mathrm{\mathbf{y}_{HA}}^H(j,k)\underline{\mathbf{C}}_\mathbf{\mathrm{v}}^{-1}(j,k)\mathrm{\mathbf{d}}(0^\circ,k)\mu_r(j,k)\\
&\quad + \mu_r^*(j,k)\mathrm{\mathbf{d}}^H(0^\circ,k)\underline{\mathbf{C}}_\mathbf{\mathrm{v}}^{-1}(j,k)\mathrm{\mathbf{y}_{HA}}(j,k)\\
&\quad - \mu_r^*(j,k)\mathrm{\mathbf{d}}^H(0^\circ,k)\underline{\mathbf{C}}_\mathbf{\mathrm{v}}^{-1}(j,k)\mathrm{\mathbf{d}}(0^\circ,k)\mu_r(j,k)\Bigr].
\end{split}
    \label{eq:joint_cond_logL}
\end{equation}

Defining $Y_\mathrm{mvdr}(j,k)\!\triangleq\!\frac{\mathrm{\mathbf{d}}^H(0^\circ,k)\underline{\mathbf{C}}_\mathbf{\mathrm{v}}^{-1}(j,k) \mathrm{\mathbf{y}_{HA}}(j,k)}{\mathrm{\mathbf{d}}^H(0^\circ,k)\underline{\mathbf{C}}_\mathbf{\mathrm{v}}^{-1}(j,k)\mathrm{\mathbf{d}}(0^\circ,k)}$, as the output of the \ac{ha} \ac{mvdr} beamformer, steered towards the front, and noting that the output noise \ac{psd} of the \ac{mvdr} beamformer is equal to $\sigma^2_\mathrm{v,mvdr}\!(l,k)\!=\!\frac{1}{\mathrm{\mathbf{d}}^H(0^\circ,k)\underline{\mathbf{C}}_\mathbf{\mathrm{v}}^{-1}(l,k)\mathrm{\mathbf{d}}(0^\circ,k)}$, and that the first three terms in (\ref{eq:joint_cond_logL}) are constant with respect to $r$, we can rewrite (\ref{eq:joint_cond_logL}) as
\begin{equation}
    \begin{split}
&\ln{p(\mathrm{\overline{\overline{\underline{\underline{\mathbf{y}}}}}_{HA}}(l)| \mathcal{H}_r(l,k),\phi_r(l,k))}\\ 
&\quad \propto \sum_{k=1}^{K}\Bigl[A_{1,r}(k)\sum\limits_{j=j_l}^{j_u}\frac{Y_\mathrm{mvdr}^*(j,k)Y_r(j,k)}{\sigma^2_\mathrm{v,mvdr}(j,k)} \\ &+ A_{1,r}^*(k)\sum\limits_{j=j_l}^{j_u}\frac{Y_r^*(j,k)Y_\mathrm{mvdr}(j,k)}{\sigma^2_\mathrm{v,mvdr}(j,k)} \\
&- |A_{1,r}(k)|^2\sum\limits_{j=j_l}^{j_u}\frac{|Y_r(j,k)|^2}{\sigma^2_\mathrm{v,mvdr}(j,k)}\Bigr].
    \end{split}
    \label{eq:joint_cond_logL_simple}
\end{equation}

By maximizing (\ref{eq:joint_cond_logL_simple}) with respect to $A_{1,r}(k)$, we can find the maximum likelihood estimate of $A_{1,r}(k)$ as
\begin{equation}
    \begin{aligned}
\hat{A}_{1,r}(k) = \frac{\sum\limits_{j=j_l}^{j_u} \frac{Y_r^*(j,k)Y_\mathrm{mvdr}(j,k)}{\sigma^2_\mathrm{v,mvdr}(j,k)}}{\sum\limits_{j=j_l}^{j_u}\frac{|Y_r(j,k)|^2}{\sigma^2_\mathrm{v,mvdr}(j,k)}}.
    \end{aligned}
    \label{eq:ml_scaleCoeff}
\end{equation}
Substituting the maximum likelihood estimate of $A_{1,r}(k)$ from (\ref{eq:ml_scaleCoeff}) in (\ref{eq:joint_cond_logL_simple}), the concentrated joint conditional log-likelihood function becomes

\begin{equation}
    \begin{split}
&\ln{p(\mathrm{\overline{\overline{\underline{\underline{\mathbf{y}}}}}_{HA}}(l)| \mathcal{H}_r(l,k), \phi_r(l,k))} \\
&\quad \propto  \sum_{k=1}^{K} \frac{\left|\sum\limits_{j=j_l}^{j_u} \frac{Y_r^*(j,k)Y_\mathrm{mvdr}(j,k)}{\sigma^2_\mathrm{v,mvdr}(j,k)}\right|^2}{\sum\limits_{j=j_l}^{j_u}\frac{|Y_r(j,k)|^2}{\sigma^2_\mathrm{v,mvdr}(j,k)}}.
   \end{split}
   \label{eq:Test_stat}
\end{equation}
We evaluate (\ref{eq:Test_stat}) for $r = 1,\hdots, R$, to solve (\ref{eq:Selection}). 
\subsection{Interpretation}\label{subsec:Interpretation}
To easily interpret the solution of the proposed method in (\ref{eq:Test_stat}), we introduce a few, additional assumptions here, which are not required to implement the proposed method in (\ref{eq:Test_stat}). Let the squared absolute correlation coefficient between the output signal of the head-steered \ac{mvdr} beamformer, $Y_\mathrm{mvdr}(j,k)$, and the $r^\mathrm{th}$ noisy \ac{rm} signal, $Y_r(j,k)$, across $D$ observations, be defined as
\begin{equation}
    \begin{aligned}
\bigl|\rho_r(l,k)\bigr|^2 \! \triangleq \! \frac{\bigl| \! \sum\limits_{j=j_l}^{j_u} \! Y_r^*(j,k)Y_\mathrm{mvdr}(j,k) \! \bigr|^2}{{\sum\limits_{j=j_l}^{j_u}\!{|Y_r(j,k)|^2}}\!{\sum\limits_{j=j_l}^{j_u}\!{|Y_\mathrm{mvdr}(j,k)|^2}}},
   \end{aligned}
   \label{eq:Test_stat_CorrelationCoeff}
\end{equation}
where $0 \! \leq \! \bigl|\rho_r(l,k)\bigr|^2 \! \leq\!1$. Let $\hat{\sigma}^2_\mathrm{y,mvdr}\!(l,k)\!=\!{\sum\limits_{j=j_l}^{j_u} \! \frac{|Y_\mathrm{mvdr}(j,k)|^2}{D}}$, denote the estimate of the output noisy \ac{psd} of the \ac{mvdr} beamformer. Assuming the output noise and the noisy \acp{psd} of the \ac{mvdr} beamformer to be constant across the $D$ observations, and using (\ref{eq:Test_stat_CorrelationCoeff}) in (\ref{eq:Test_stat}), we can rewrite (\ref{eq:Selection}) as

\begin{equation}
    \begin{aligned}
    \hat{r}_{\mathrm{prop}}(l) = \operatorname*{argmax}_{r \in \{1, \hdots, R\}} \sum_{k=1}^{K}\frac{\hat{\sigma}^2_\mathrm{y,mvdr}(l,k)}{\sigma^2_\mathrm{v,mvdr}(l,k)} \bigl|\rho_r(l,k)\bigr|^2.
    \end{aligned}
    \label{eq:Selection_final}
\end{equation}

From (\ref{eq:Selection_final}), it is clear that the proposed \ac{cs} method selects the \ac{rm} channel, which leads to the highest weighted squared absolute correlation coefficient between the output signal of the head-steered \ac{mvdr} beamformer, $Y_\mathrm{mvdr}(l,k)$, and the $r^\mathrm{th}$ \ac{rm} signal, $Y_r(l,k)$. Note that the weighting across the $k$ frequency bins is equal to the estimated posterior \ac{snr} of the \ac{ha} \ac{mvdr} beamformer, $\textstyle \hat{\gamma}_r(l,k)\!\triangleq\!\frac{\hat{\sigma}^2_\mathrm{y,mvdr}(l,k)}{\sigma^2_\mathrm{v,mvdr}(l,k)}$. When the output \ac{snr} of the \ac{ha} \ac{mvdr} beamformer is low, $\textstyle \hat{\gamma}_r(l,k)$ tends to $1$, and when it is high, $\textstyle \gamma_r(l,k)$ is greater than $1$.  
\subsection{Implementation}\label{subsec:Close_Implementation}
To implement the \ac{mvdr} beamformer in (\ref{eq:Test_stat}), we require an estimate of the \ac{ha} noise \ac{cpsdm}, $\underline{\mathbf{C}}_\mathbf{\mathrm{v}}(l,k)$, which typically, would require a reliable \ac{vad} corresponding to the target talker \cite{breithaupt2008noise}. In low \ac{snr} situations, with multiple competing talkers, such a target \ac{vad} is hard to realize. To circumvent this challenge, and due to the equivalence of the \ac{mvdr} to the \ac{mpdr} beamformer \cite{van2002optimum}, we use the output of the \ac{mpdr} beamformer, $Y_\mathrm{mpdr}(j,k)\!\triangleq\!\frac{\mathrm{\mathbf{d}}^H(0^\circ,k)\underline{\mathbf{C}}_\mathbf{\mathrm{y}}^{-1}(j,k) \mathrm{\mathbf{y}_{HA}}(j,k)}{\mathrm{\mathbf{d}}^H(0^\circ,k)\underline{\mathbf{C}}_\mathbf{\mathrm{y}}^{-1}(j,k)\mathrm{\mathbf{d}}(0^\circ,k)}$, in (\ref{eq:Test_stat}) as it only requires the noisy \ac{cpsdm} of the \ac{ha} signals, $\underline{\mathbf{C}}_\mathbf{\mathrm{y}}(l,k)$. In acoustic beamformer applications, the \ac{mvdr} beamformer is often preferred over the \ac{mpdr} beamformer, as the latter tends to introduce audible distortions if the \ac{ratf}, $\mathbf{d}(0^\circ,k)$ is not estimated accurately. However, for our purposes, the output of the \ac{mpdr} beamformer is not for listening, but for choosing the correct \ac{rm}, which makes the use of the \ac{mpdr} beamformer preferable in this situation. Furthermore, the solution in (\ref{eq:Test_stat}) requires the noise and noisy output \acp{psd} of the \ac{mvdr} beamformer, $\textstyle \sigma^2_\mathrm{v,mvdr}(l,k)$, $\textstyle \hat{\sigma}^2_\mathrm{y,mvdr}(l,k)$, respectively, which are equal to the corresponding output \acp{psd} of the \ac{mpdr} beamformer, $\textstyle \sigma^2_\mathrm{v,mpdr}(l,k)$, $\textstyle \hat{\sigma}^2_\mathrm{y,mpdr}(l,k)$ \cite{van2002optimum}. As above, to estimate the noise \ac{psd} of the \ac{mpdr} beamformer, $\sigma^2_\mathrm{v,mpdr}(l,k)$, requires a good target talker \ac{vad}, which is hard to realize in the presence of multiple competing talkers. Therefore, for implementation purposes, we use $\sigma^2_\mathrm{v,mpdr}(l,k)=\hat{\sigma}^2_\mathrm{y,mpdr}(l,k)$, such that every frequency bin in (\ref{eq:Selection_final}) is weighted equally. Note that this approximation is reasonable for low \acp{snr}. Furthermore, as we demonstrate in Section \ref{subsec:Close_Results}, using $\hat{\sigma}^2_\mathrm{y,mpdr}(l,k)$ in place of $\sigma^2_\mathrm{v,mpdr}(l,k)$, leads to very similar \ac{cs} performance for realistic \acs{snr} in acoustic scenes.

We demonstrate the proposed \ac{cs} method using close-talking \acp{rm} and table microphone arrays in Section \ref{sec:Close_RM}. The signals used in the simulations are sampled at a sampling frequency of $f_s = 16$ kHz and processed in the \ac{stft} domain, using $32$ ms square-root Hann analysis and synthesis windows, with $50\%$ overlap and FFT size of $N_\mathrm{FFT} = 512$ samples. This window length is chosen to ensure that it is longer than the effective duration of all acoustic impulse responses from source to microphones used. As the output of the beamformer is used only for the \ac{cs}, which is not a time critical operation (\ref{eq:Test_stat}), the use of longer analysis frames for \ac{stft} processing is permitted in the present context. This, consequently, increases the frequency resolution of processing and will accommodate longer acoustic impulse responses between the the different candidate talkers and the \ac{ha} microphones, thereby eliminating the need for correlating across different time lags, as typically done when using shorter analysis frames. The duration of all signals used in the simulations is $30$ s. The noisy \ac{cpsdm} of the \ac{ha} microphone signals, $\underline{\mathbf{C}}_\mathbf{\mathrm{y}}(l,k)$ is estimated recursively at every frequency bin, $k$, as
\begin{equation}
    \begin{split}
        \underline{\mathbf{C}}_\mathbf{\mathrm{y}}(l,k)&= \alpha_y\underline{\mathbf{C}}_\mathbf{\mathrm{y}} (l -1,k)\\ &+ (1 - \alpha_y)  \left[\mathbf{y}_\mathrm{HA}(l,k) \mathbf{y}_\mathrm{HA}^H\!(l,k)\right],
    \end{split}
    \label{eq:C_y_recursive}
\end{equation}
where the smoothing co-efficient is $\alpha_y  \!= \!0.8465$, corresponding to a time-constant of $96$ ms. We take the left-front \ac{ha} microphone to be the \ac{ha} reference microphone. For the simulations, we use only the left bilateral beamformer using the $M\!=\!4$ \ac{ha} microphones, to implement the proposed method.
\section{Performance analysis}\label{sec:Close_RM}
We analyze the performance of the proposed \ac{cs} method in multi-talker scenarios, using close-talking \acp{rm} in Section \ref{subsec:Close_RM} and using table microphone arrays in Section \ref{subsec:Beam_application}.
\subsection{Simulation experiments using close-talking RMs}\label{subsec:Close_RM}
One of the common challenging situations for adults using \acp{ha} would be the dinner table settings, while for children using \acp{ha} it would be the classroom setting. Currently there are multiple wireless \ac{ha} accessories like clip-on microphones or personal microphones that can be worn by the candidate talkers, and the signals are transmitted from a single or multiple \acp{rm} to the \ac{ha} directly. Therefore, we evaluate the performance of the proposed \ac{cs} method for the situation where each remote channel is captured by a \ac{rm}, e.g., clip-on microphones. 
\subsubsection{Simulation setup}\label{subsec:Simulated_rir}
\begin{figure}[htbp]
    \centering
    \includegraphics[width=0.8\linewidth]{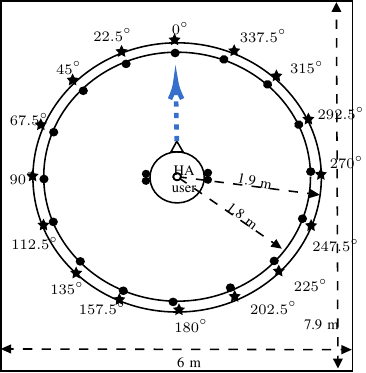}
    \caption{Schematic diagram showing the source position `\protect\showpgfstar', and the chosen \ac{rm} positions `\protect\showpgfcircle',  for the synthetic microphone \ac{rir} generated, based on the room characteristics in \cite{moore_personalized_2019}.\label{fig:SytheticData_Model}}
\end{figure}

We simulate the acoustic scene shown in Figure \ref{fig:SytheticData_Model}, with a single \ac{ha} user, wearing bilateral head-mounted \acp{ha}, i.e., \ac{ha} units on both ears, in the presence of $N\!+\!1$ candidate talkers, each wearing a \ac{rm}, placed close to their mouths. The target talker is located in the head-steered direction of the \ac{ha} user, i.e.,  $\theta_1 = 0^\circ$, while the remaining $N$ candidate talkers are competing talkers. We use speech signals from the English conversation dataset in \cite{sorensen2018task}, which consists of multiple two person conversation signals. We randomly assign the signals to all the candidate talkers. As shown in Figure \ref{fig:SytheticData_Model}, we place a \ac{rm} close to each candidate talker, i.e., $R\!=\!N\!+\!1$ \acp{rm}, and every \ac{rm} captures the target talker signal, but also the competing speech signals. In addition to the $R$ \acp{rm}, we consider $M\!=\!4$ \ac{ha} microphones of the bilateral \acp{ha} worn by the user, with $\frac{M}{2}\!=\!2$ microphones on each ear. We arbitrarily chose the left front microphone to be the \ac{ha} reference microphone.

The \ac{ha} microphone signals are simulated by convolving each candidate talker signal with the \ac{hahrir} from the respective candidate talker to the \ac{ha} microphones, and by summing the convolved signals. Similarly, the \ac{rm} signals are simulated by convolving the candidate talker signals with the \ac{rir} from each candidate talker to the \acp{rm} in question, and summing the convolved signals. The \acp{hahrir}, measured on a \ac{hats} mannequin (\textit{Subject $28$}) in \cite{moore_personalized_2019}, are used to simulate the \ac{ha} microphone signals, while the \ac{rir} with respect to the close-talking \acp{rm} are generated using the image method proposed in \cite{allen1979image}, using the room characteristics described in \cite{moore_personalized_2019}. We simulate the \ac{ha} user to be located at the center of the room and the candidate talkers to be located on a circle of radius $1.9$ m, at $16$ azimuthal angles, with the \acp{rm} located $0.2$ m in front of the candidate talkers' mouth, see Figure \ref{fig:SytheticData_Model}. We simulate the background noise at each \ac{ha} microphone to be isotropic \ac{ssn} such that \ac{snr} of the target talker signal with respect to the background noise is $15$ dB at the reference microphone, but note that the \ac{snr} reduces with every competing talker added to the scene. The isotropic \ac{ssn} is generated by convolving \ac{ssn}, whose spectral shape is determined from the long-term spectrum of the speech signals in \cite{garofolo1993darpa}, with the \acp{hahrir} from the $16$ azimuthal source positions to each \ac{ha} microphone, and summing them at each microphone. We use independent \ac{ssn} signals as the background noise at \acp{rm}, added at the same energy level as the isotropic \ac{ssn} signal at the \ac{ha} reference microphone. The frontal-\ac{ratf}, $\mathbf{d}(0^\circ,k)$, required in (\ref{eq:joint_cond_logL}) is taken from \cite{moore_personalized_2019}.

The acoustic setup considered in Figure \ref{fig:SytheticData_Model}, there are $15$ possible competing talker positions. In the following analysis we compare the \ac{cs} performance by averaging the results across $40$ combinations of uniformly randomly selected competing talker locations for $N\!=\!\{2,\hdots,8\}$, competing talkers. Furthermore, to assess the rapidity of the proposed method in adapting to the dynamics of the acoustic environment, we analyze the performance of \ac{cs} as a function of the integration time, $T_\mathrm{int}$, which corresponds to the number of $D$ observations in (\ref{eq:multi_observations_HA}), used to correlate the output signal of the \ac{mpdr}, $Y_\mathrm{mpdr}(l,k)$ and the \ac{rm} signals, $Y_r(l,k)$ in (\ref{eq:Selection_final}). 

\begin{figure}[htbp]
    \centering
    \begin{minipage}[t]{0.99\linewidth}
        \centering
        \includegraphics[width=0.4\linewidth]{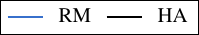}
    \end{minipage}
    \begin{minipage}[t]{0.99\linewidth}
        \centering
        \includegraphics[width=0.85\linewidth]{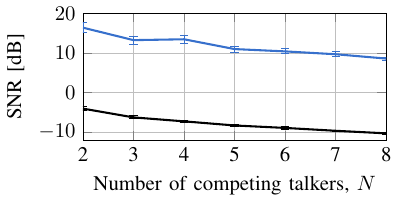}
    \end{minipage}
    \caption{Input SNR at the \ac{ha} reference microphone and the \ac{rm} placed on the target talker at $\theta_1 = 0^\circ$, across $N = \{2,\hdots, 8\}$, number of competing talkers, for the acoustic scene in Figure \ref{fig:SytheticData_Model}.}
    \label{fig:micSelect_imagesource_SNR}
\end{figure}

In Figure \ref{fig:SytheticData_Model}, the target remote channel is the signal captured by the \ac{rm} that is placed closest to the target talker located in the head-steered direction of the \ac{ha} user, i.e.,  $\theta_1 = 0^\circ$. As mentioned previously, the \ac{snr} at the \ac{ha} reference microphone and the \acp{rm} lowers with every competing talker added to the scene. Figure \ref{fig:micSelect_imagesource_SNR} shows the average input \ac{snr} at the \ac{ha} reference microphone and the average input \ac{snr} at the \ac{rm} placed on the target talker, across the number of competing talkers, $N = \{2,\hdots,8\}$, averaged across the $40$ combinations of the competing talker positions, where the standard deviation is shown as error bar plots.

\subsubsection{Baseline methods}\label{subsec:SOTA}
When each candidate talker is equipped with a \ac{rm}, we can use the \textit{turn-taking} behaviour \cite{sacks1978simplest,mccowan2003modeling}, between the \ac{ha} user and the candidate talkers to find the target talker \ac{rm} channel. For the acoustic scene described in Section \ref{subsec:Simulated_rir}, we compare the performance of the proposed \ac{cs} method, to the \textit{turn-taking} based methods, \ac{ncc} \cite{harma2009conversation} and \ac{mog} \cite{hoang2022minimum}, in selecting the target talker channel. \newline
%
\noindent a) Normalized Cross-Correlation (\ac{ncc}) \cite{harma2009conversation}: The \ac{ncc} method identifies the target talker as the candidate talker whose binary voice activity sequence complements the \ac{ha} user's own-voice binary voice activity sequence. If $V_0(l)$ denotes the \ac{ha} user's own-voice voice activity sequence, and $V_r(l)$ denotes the voice activity sequence of the $r^\mathrm{th}$ candidate talker, at the $l^\mathrm{th}$ time-frame, the \ac{ncc} decision can be written as  \cite{harma2009conversation}
\begin{equation}
    \begin{aligned}
        \hat{r}_\mathrm{NCC}(l) = \operatorname*{argmax}_{r \in \{1, \hdots, R\}} \frac{1-\min_{p\in\{n_l, n_u\}}\mathcal{R}_{0,r}(p)}{2},
    \end{aligned}
    \label{eq:NCC_decision}
\end{equation}
where $\mathcal{R}_{0,r}(p)$ is the cross-correlation, computed using an integration time, $T_\mathrm{int}$, between the \ac{ha} user's own-voice voice activity sequence, $V_0(l)$, and the $r^\mathrm{th}$ candidate voice activity sequence, $V_r(l)$, at lag $p$.\newline
%

\noindent b) Minimum Overlap-Gap (\ac{mog}) \cite{hoang2022minimum}: The \ac{mog} algorithm \cite{hoang2022minimum}, identifies the target talker as the candidate talker, whose binary voice activity sequence, $V_r(l)$, maximizes the \ac{mse} with respect to the \ac{ha} user's own-voice binary voice activity sequence, $V_0(l)$ \cite{hoang2022minimum}:
\begin{equation}
    \begin{aligned}
        \hat{r}_\mathrm{MOG}(l) = \operatorname*{argmax}_{r \in \{1, \hdots, R\}} \mathbb{E}\left[\left(V_0(l)-V_r(l)\right)^2\right],
    \end{aligned}
    \label{eq:MOG_decision}
\end{equation}
computed using an integration time, $T_\mathrm{int}$, at time-frame, $l$.
%
\subsubsection{Results}\label{subsec:Close_Results}
\begin{figure*}[htbp]
    \centering
    \begin{minipage}[t]{0.99\textwidth}
    \centering
        \includegraphics[width=0.75\linewidth]{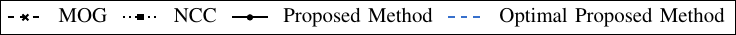}
    \end{minipage}
    \begin{minipage}[t]{0.99\textwidth}
    \centering
    \includegraphics[width=0.98\linewidth]{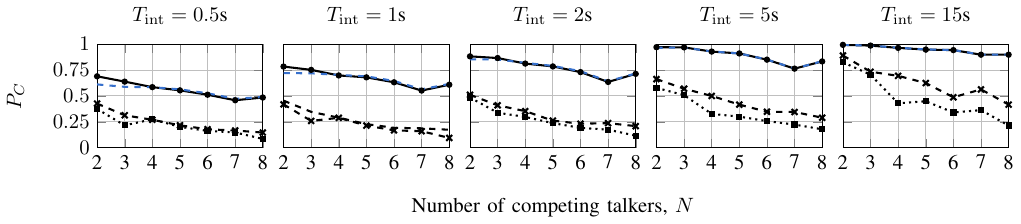}
    \end{minipage}
    \caption{Probability of correct target talker selection, $P_C$, using \ac{ncc}, \ac{mog} and the Proposed Method, and Optimal Proposed Method, in the presence of $N=\{2,3,4,5,6,7,8\}$ competing talkers, for $T_\mathrm{int} = \{0.5,1,2,5,15\}s$}.
\label{fig:Compare_MOG_NCC_Pinocchio_Oracle}
\end{figure*}
We compare the proposed \ac{cs} method against \ac{ncc} and \ac{mog}, in a multi-talker scenario. In our implementation of \Ac{ncc} and \ac{mog}, we use oracle \acp{vad} $V_r(l,k)$, $V_0(l,k)$, estimated from the $R$ clean candidate talker signals and \ac{ha} user's own-voice signal, respectively. This gives \ac{ncc} and \ac{mog} a very significant advantage over the proposed \ac{cs} method. 

In the following analysis, we introduce the probability of correct selection, $P_C$, of identifying the correct \ac{rm} amongst the candidate \acp{rm}. $P_C$ is defined as
\begin{equation}
    \begin{aligned}
        P_C \triangleq \frac{\text{No. of time-frames the target \ac{rm} is selected}}{\text{Total no. of time-frames}}.
    \end{aligned}
    \label{eq:def_Pc}
\end{equation}
Intuitively, when $P_C = 1$, it implies that the correct \ac{rm} is chosen at all time-frames, while $P_C = 1/R$ is a performance lower bound, which corresponds to the random selection of the target remote channel.

Figure \ref{fig:Compare_MOG_NCC_Pinocchio_Oracle} shows the performance of the proposed \ac{cs} method, \ac{ncc} and \ac{mog} in terms of $P_C$, as a function of number of competing talkers, $N \!= \!\{2,\hdots,8\}$, for different integration times, $T_\mathrm{int} \!= \! \{0.5, 1, 2, 5, 15 \}$s. As expected, the proposed method is more accurate for fewer competing talkers and longer integration times, $T_\mathrm{int}$, because, with longer integration times, $T_\mathrm{int}$, the correlation estimated between the output of the \ac{mpdr} and the \ac{rm} channel in (\ref{eq:Test_stat}) stabilizes. Note that since \ac{ncc} and \ac{mog} rely on \textit{turn-taking}, they require longer integration times, $T_\mathrm{int}$, to identify the target talker accurately. From Figure \ref{fig:Compare_MOG_NCC_Pinocchio_Oracle}, it is clear that the proposed method outperforms \ac{ncc} and \ac{mog} in selecting the target talker \ac{rm}, under all the number of competing talkers, $N$, and integration times, $T_\mathrm{int}$, considered, despite \ac{mog} and \ac{ncc} operating with ideal \acp{vad}.

To study the impact of the approximation, $\sigma_\mathrm{v,mpdr}^2(k,l)\!\approx\!\hat{\sigma}_\mathrm{y,mpdr}^2(k,l)$, that is used to implement the `Proposed method', the `Optimal Proposed method' curves in Figure \ref{fig:Compare_MOG_NCC_Pinocchio_Oracle} show the performance using $\sigma_\mathrm{v,mpdr}^2(k,l)$, computed with access to noise components of the \ac{ha} microphone signals in isolation, i.e., oracle information. As expected in Sec. \ref{subsec:Close_Implementation}, at realistic \acp{snr} used in the acoustic scene, see Figure \ref{fig:micSelect_imagesource_SNR}, this approximation remains valid and the performance of the methods is very similar.

From Figure \ref{fig:Compare_MOG_NCC_Pinocchio_Oracle}, it is evident that for all methods compared, performance increases with integration time, $T_\mathrm{int}$. While the proposed \ac{cs} method benefits from longer integration time in order to correlate the output of the \ac{mpdr} with the \ac{rm} signals in (\ref{eq:Test_stat}), for the baseline \textit{turn-taking} methods, longer integration time is useful to acquire more information regarding the speech gaps and overlaps during a conversation between the \ac{ha} user and the candidate talkers. Nevertheless, the proposed \ac{cs} method outperforms the baseline methods, even during shorter integration times.
\subsubsection{Robustness to RATF mismatch}\label{subsec:RATF_analysis}
\begin{figure}[htbp]
    \centering
    \begin{minipage}[t]{0.99\linewidth}
        \centering
        \includegraphics[width=0.75\linewidth]{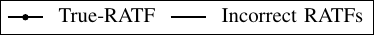}
    \end{minipage}
    \begin{minipage}[t]{0.99\linewidth}
        \centering
        \includegraphics[width=0.99\linewidth]{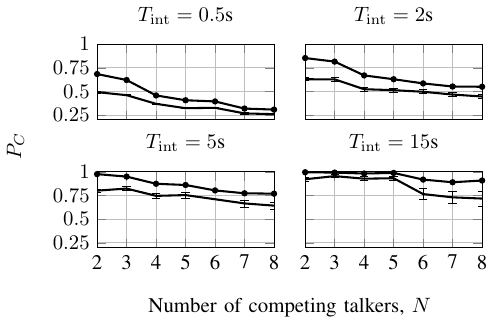}
    \end{minipage}
    \caption{Probability of correct target talker selection, $P_C$, using True-\ac{ratf}, \ac{ratf} vectors from different subjects in \cite{moore_personalized_2019}, in the presence of $N=\{2,\hdots,8\}$ competing talkers, for $T_\mathrm{int} = \{0.5,2,5,15\}$s.}
    \label{fig:Compare_RTF_20Subejcts}
\end{figure}
The proposed \ac{cs} method assumes that the target talker is located along the head-steering direction of the \ac{ha} user and requires the frontal \ac{ratf} vectors, $\mathbf{d}(0^\circ,k)$, in (\ref{eq:joint_cond_logL}). In Section \ref{subsec:Close_Results}, we used the \ac{ratf} vectors, measured on the \ac{hats} mannequin, when simulating the acoustic scene. In reality, the frontal \ac{ratf} would vary depending on the head size/shape of the \ac{ha} user, head-and-torso acoustics and the position of the \acp{ha}. Nevertheless, the matched \ac{ratf} situation simulated in Sec. \ref{subsec:Close_Results} could be achieved even in real-life situations using \textit{in-situ} calibration methods \cite{jensen2017self} for estimating user-dependent individualized \ac{ratf} vectors. 

Even so, it is of interest to understand the robustness of the proposed \ac{cs} method to incorrect frontal \ac{ratf} vectors used. To do this, we test the proposed method using frontal \ac{ratf} vectors measured on different heads, from \cite{moore_personalized_2019}, where the authors measured the \acp{hahrir} on $4$ HATS mannequins, and $41$ human subjects, using the same \ac{ha} device. In this assessment, in addition to the frontal \ac{ratf} vector used in Section \ref{subsec:Simulated_rir} (\textit{Subject} 28 in \cite{moore_personalized_2019}), we use the frontal \ac{ratf} from $20$ randomly selected subjects. We compare the probability of correct selection, $P_C$ (\ref{eq:def_Pc}), of the target \ac{rm} located along the head-steering direction, when using `True-\ac{ratf}'(frontal \ac{ratf} of \textit{Subject} 28), and `Incorrect-\acp{ratf}', (frontal \acp{ratf} from the $20$ subjects). 

Figure \ref{fig:Compare_RTF_20Subejcts} shows the performance of the proposed \ac{cs} method in terms of $P_C$, using the `True-\ac{ratf}' and the `Incorrect-\acp{ratf}', across the number of competing talkers, $N\!=\!\{2,\hdots,8\}$, for integration times, $T_\mathrm{int} \!=\! \{0.5,2,5,15\}$s. The results from the $20$ subjects of `Incorrect-\acp{ratf}' are averaged and their standard deviations are shown as error bars. Clearly, using the `True-\ac{ratf}' results in better performance than using the `Incorrect-\acp{ratf}'. This is because, the spatial nulls introduced by the \ac{ha} \ac{mpdr} beamformer are less deep using the `Incorrect-\acp{ratf}' than using the `True-\ac{ratf}'. This results in lower noise reduction performance of the head-steered \ac{mpdr} beamformer, which reduces, $P_C$. Nevertheless, when using higher integration times, $T_\mathrm{int}$, the performance loss from using the `Incorrect-\acp{ratf}' reduces. At lower integration times, the proposed \ac{cs} method outperforms the baselines (see Figure \ref{fig:Compare_MOG_NCC_Pinocchio_Oracle}), even with incorrect frontal \acp{ratf} used.
\subsubsection{Robustness to non-frontal target talker locations}\label{subsec:NonFrontal_RATF_analysis}
In dynamic acoustic situations, it is quite likely for the \ac{ha} user to be moving their head to listen to multiple candidate talkers in the scene, and the \ac{ha} user may therefore not perfectly steer their head towards the target talker at $ \theta_1 = 0^\circ$, as assumed in (\ref{eq:Hypotheses}). Alternatively, the target talker may not be located exactly in front of the \ac{ha} user and the \ac{ha} user's head might occasionally be steered away from the target talker, while their eye gaze is steered towards the target talker. When the target talker location is non-frontal, i.e., $\theta_1 \neq 0^\circ$, the assumption in (\ref{eq:Hypotheses}) is violated, and therefore the performance is expected to be lower compared to when the assumption of frontal target talker, i.e., $\theta_1\,=\, 0 ^\circ$ is met. In such cases, additional methods, e.g., \textit{turn-taking} methods \cite{harma2009conversation,hoang2022minimum}, could be employed to enhance performance and improve overall results. Alternatively, or additionally, more advanced tracking methods might be employed which make better use of past values of the log-likelihood function in (\ref{eq:Test_stat}) by implicitly or explicitly taking into account the dynamics of head movements during conversations. Such methods are well outside the scope of this paper. However, for the purpose of a comprehensive analysis of the proposed \ac{cs} method, this section presents the performance of the proposed \ac{cs} method for non-frontal target talker locations.  

To do so, for the acoustic scene in Sec. \ref{subsec:Close_Results}, we test the proposed \ac{cs} method using non-frontal target locations, $\theta_1 = \{22.5^\circ, 337.5^\circ, 45^\circ, 315^\circ\}$, while using a frontal \ac{ratf}, $\mathbf{d}(0^\circ,k)$, from \cite{moore_personalized_2019}, in (\ref{eq:joint_cond_logL}). We compare the performance of the proposed \ac{cs} method for a frontal target talker, $\theta_1\,=\, 0^\circ$, against non-frontal target talker locations, $\theta_1  \neq 0^\circ$. The performance is estimated in terms of probability of correct selection, $P_C$ (\ref{eq:def_Pc}), of the target \ac{rm} located closest to the target talker located at $\theta_1$.
\begin{figure}[htbp]
    \centering
    \begin{minipage}[t]{0.99\linewidth}
        \centering
        \includegraphics[width=0.9\linewidth]{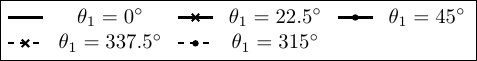}
    \end{minipage}
    \begin{minipage}[t]{0.99\linewidth}
        \centering
        \includegraphics[width=0.8\linewidth]{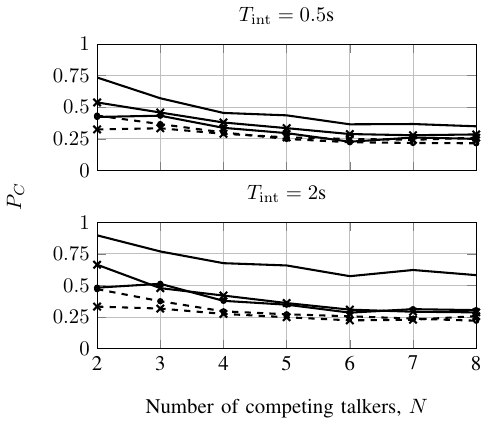}
    \end{minipage}
    \caption{Probability of correct target talker selection, $P_C$, using \ac{ratf} vectors, $\mathbf{d}(\theta,k)$ steered towards $\theta = \{0^\circ,22.5^\circ,337.5^\circ,45^\circ,315^\circ\}$ in \cite{moore_personalized_2019}, in the presence of $N=\{2,\hdots,8\}$ competing talkers, for $T_\mathrm{int} = \{0.5,2\}$s.}
    \label{fig:Compare_RTF_DOA_N_rerun}
\end{figure}

Figure \ref{fig:Compare_RTF_DOA_N_rerun} shows the performance of the proposed \ac{cs} method in terms of $P_C$, for target talker locations, $\theta_1 = \{0^\circ, 22.5^\circ, 337.5^\circ, 45^\circ, 315^\circ\}$, across the number of competing talkers, $N = \{2,3,4,5,6,7,8\}$, for the integration times $T_\mathrm{int} = \{0.5, 2\}$ s. In the simulations, we ensure that a competing talker is not located along the head steering direction. The results shown are averaged over $40$ combinations of the competing talker positions.

From Figure \ref{fig:Compare_RTF_DOA_N_rerun}, it can be observed that the proposed \ac{cs} method performs best when the \ac{ha} user's head is steered towards the direction of the target talker ($\theta_1 = 0^\circ$), as assumed in (\ref{eq:Hypotheses}). Moreover, as expected, the loss in probability of correct selection, $P_C$, increases with the angle mismatch. Note that, we use the left front microphone as the reference microphone for the \ac{ha} \ac{mpdr}. Due to the head shadow effect, the spatial response of the \ac{ha} \ac{mpdr} becomes asymmetric for target locations on the right of the \ac{ha} user, i.e., $180^\circ<\theta_1<360^\circ $, compared to targets located to the left, i.e., $0^\circ<\theta_1<180^\circ $. Therefore, in Figure \ref{fig:Compare_RTF_DOA_N_rerun}, we observe $P_C$ to be lower for target locations on the right at $\theta_1 = \{315^\circ , 337.5^\circ\}$  when compared to target locations on the left at $\theta_1 = \{22.5^\circ , 45^\circ\}$, even-though the are symmetrically located w.r.t the \ac{ha} user's head. In such situations, the loss in performance could be diminished by combining the left-right \ac{ha} channel selection decisions. Nevertheless, the performance of the proposed \ac{cs} method even for non-frontal target talker locations exceeds the performance of the baseline methods discussed previously in Section \ref{subsec:Close_Results} (cf. Figure \ref{fig:Compare_MOG_NCC_Pinocchio_Oracle}). Moreover, similar to the trend observed in Sec. \ref{subsec:Close_Results}, in the presence of fewer competing talkers, e.g., $N<4$, $P_C$ increases with longer integration times, $T_\mathrm{int}$. 
\subsection{Simulation experiments using table microphone arrays}\label{subsec:Beam_application}
\begin{figure}[htp]
    \centering
    \includegraphics[width=0.9\linewidth,height=0.7\linewidth]{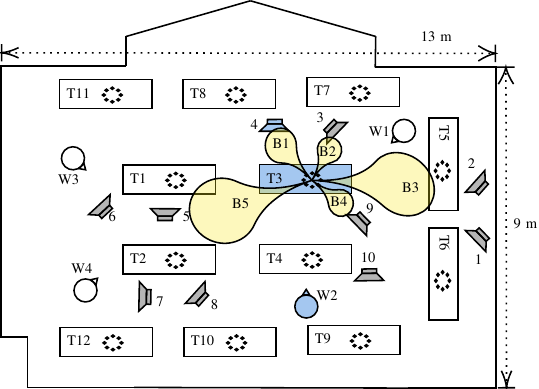}
    \caption{Sketch showing the target talker position `4', the table microphone array, `T3', and the position of \ac{ha} user, `W2', as measured in \cite{corey2019massive}. Also, the \ac{mpdr} beamformers, B1-B5, from the table microphone T3 are shown.}
    \label{fig:BeamSelect_Model}
\end{figure}%
Since using individual \acp{rm} for each candidate talker is not always feasible, here we consider a situation where a table microphone is accessible to the \ac{ha} user. In this situation, the table microphone separates the multiple candidate talkers in the scene, e.g., via beamforming or using speaker separation algorithms, among which the proposed \ac{cs} method must select the target talker. In particular, we demonstrate that the proposed \ac{cs} method accurately selects the channel containing the frontal target talker, when choosing among enhanced candidate talker signals from beams steered in different directions in the room.
\begin{figure*}[htbp]
    \centering
    \begin{minipage}[t]{0.99\textwidth}
    \centering
        \includegraphics[width=0.75\linewidth]{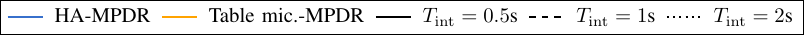}
    \end{minipage}
    \begin{minipage}[t]{0.99\textwidth}
    \centering
    \includegraphics[width=0.99\linewidth]{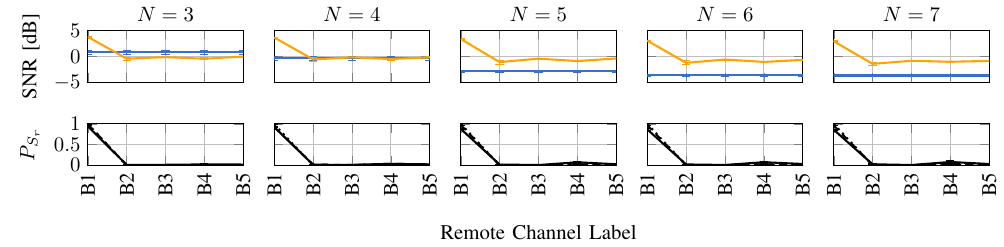}
    \end{minipage}
    \caption{Performance of the proposed \ac{cs} method in terms of $P_{S_r}$ for beams:$\{$B$1$, B$2$, B$3$, B$4$, B$5\}$, in the presence of $N=\{3,4,5,6,7\}$, using integration times, $T_\mathrm{int}= \{0.5, 1, 2\}$s for table microphone arrays.}
\label{fig:BeamSelect_Compare}
\end{figure*}%
\subsubsection{Simulation Setup}
To simulate the acoustic scene, we use \acp{hahrir} and \acp{rir} from \cite{corey2019massive}, measured in a large, reverberant conference room ($T_{60} = 800$ ms). The acoustic scene described in \cite{corey2019massive}, is illustrated in Figure \ref{fig:BeamSelect_Model}, where four \ac{ha} users are denoted by W1-W4, $10$ loudspeakers are denoted by numbers $1\!-\!10$, that act as candidate sources. Multiple tables are placed in the room, with $12$ table microphone arrays, denoted by T1-T12, each containing $8$ microphones in a circular array. We use the \ac{ha} user, W2, with $M\!=\!4$ microphones (front and rear microphones on both ears), the table microphone array, T3, with $8$ microphones, the loudspeaker located at `4', as the target talker and the remaining loudspeakers as competing talkers. The head-steered \ac{mpdr} beamformer of the \ac{ha} user, W2, is pointing frontal towards the target talker located at `4'. The table microphone, T3, has fixed \ac{mpdr} beamformers directed towards candidate talkers: $\{4,3,2,9,5\}$, denoted by, B1-B5 (see Figure \ref{fig:BeamSelect_Model}). The output of the table microphone \ac{mpdr} beamformers represent the $R$ remote channels, from which the proposed method must identify the channel that best corresponds to the signal from the frontal talker at `4'. Note that the table microphone beamformers are chosen to be \ac{mpdr} beamformers for convenience, however, in reality can represent other speech enhancement algorithms. 

The microphone signals at the \ac{ha}, W2, are generated by convolving the candidate talker signals with \ac{hahrir} from the corresponding candidate talker to the \ac{ha} microphones. Similarly, the signals at the table microphone array, T3, are generated by convolving the candidate talker signals with \acp{rir} from the corresponding candidate talker to the \acp{rm}. We add the background noise recorded in \cite{corey2019massive}. We implement the \ac{mpdr} beamformers at the table microphone, T3, steered towards the candidate talkers: $\{4,3,2,9,5\}$, using the \acp{atf} from \cite{corey2019massive}. For the target talker at `4', the channel of interest is the output of \ac{mpdr} beamformer, B1. 
\subsubsection{Results}
Our goal is to identify how accurately the proposed method identifies the target \ac{rm} (many speech enhancement systems can be envisaged from this, but they are beyond our scope). Hence, we quantify the performance of the proposed method using the probability of selection of the $r^\mathrm{th}$ channel,
\begin{equation}
    \begin{aligned}
         P_{\!S_r} \!=\! \frac{\text{  No. of time-frames $r^\mathrm{th}$ channel is selected}}{\text{Total no. of time-frames}},
    \end{aligned}
    \label{eq:def_Ps_r}
\end{equation}
where $r\! \in\! \{1,\hdots, R\}$.

The first row of Figure \ref{fig:BeamSelect_Compare} shows the output \acp{snr} of the \ac{ha} \ac{mpdr} beamformer and the fixed \ac{mpdr} beamformers, B1-B5, at the table microphones, T3, where the \ac{snr} is defined as the power of the target talker signal relative to the power of the noise signals in the output signal of the \ac{mpdr} beamformer. There are $9$ competing talker locations, and the output \acp{snr} plotted are averaged across $20$ combinations of uniformly randomly selected competing talker locations for $N\!=\!\{3,\hdots,7\}$, and the standard deviations are shown as error bars. From the \ac{snr} plots, we infer that it would be highly beneficial to be able to select the channel corresponding to the output of the \ac{mpdr} beamformer, B1, as it has an \ac{snr} that is $3\!-\!6$ dB higher than any other \ac{mpdr} beamformer, including that of the \ac{ha} \ac{mpdr} beamformer, which serves as a natural baseline. 

The second row of Figure \ref{fig:BeamSelect_Compare} shows the results in terms of $P_{S_r}$, for competing talkers, $N \! = \! \{3,\hdots, 7\}$, for different integration times, $T_\mathrm{int} \!=\! \{0.5, 1, 2\}$s, across $20$ combinations of competing talkers positions. As expected, $P_{S_r}$ of the channel from B1, is high, i.e., $P_{S_r}\geq 0.9$ for B1 for all number of competing talkers, $N$ and integration times, $T_\mathrm{int}$.

\section{Conclusion}\label{sec:Conclusion}
We consider the problem of \acf{cs} for \acf{ha} applications using \acfp{rm}, in acoustic situations with multiple competing talkers. We use the head-steering direction of the \ac{ha} user to pose the \ac{cs} problem in a multiple hypothesis test framework and derive a maximum likelihood solution. The proposed method chooses the \ac{rm} channel that leads to the highest weighted squared absolute correlation coefficient between the output of the head-steered \ac{ha} \acf{mvdr} beamformer and the \ac{rm} channel. Through simulations using realistic acoustic scenes using close-talking \acp{rm} and table microphone arrays, we show that the proposed \ac{cs} method accurately selects the \ac{rm} channel located in the head-steering direction of the \ac{ha} user. Furthermore, we demonstrate that the proposed \ac{cs} method, without the need for additional sensors, significantly outperforms its baselines in the presence of multiple competing talkers.

\newpage

\section{Biography Section}
\vskip -2.5\baselineskip plus -1fil
\begin{IEEEbiographynophoto}
{VASUDHA SATHYAPRIYAN} received the M.Sc. degree (cum laude) in Electrical Engineering from Delft University of Technology, Delft, The Netherlands in 2020. She is currently pursuing a Ph.D degree in Electrical Engineering from Aalborg University, in collaboration with Demant A/S, Denmark.
\end{IEEEbiographynophoto}
\vskip -2.5\baselineskip plus -1fil
\begin{IEEEbiographynophoto}
{MICHAEL S. PEDERSEN} received the M.S. degree from the Technical University of Denmark (DTU), Lyngby, Denmark, in 2003 and the Ph.D. degree from the Section for Intelligent Signal Processing, Department of Mathematical Modelling, DTU, in 2006. Since 2001, he has been with the hearing instrument company, Demant A/S, as a Principal Specialist. He has authored or coauthored more than 20 peer reviewed publications, and he is listed as an inventor on more than 60 patent applications.
\end{IEEEbiographynophoto}
\vskip -2.5\baselineskip plus -1fil
\begin{IEEEbiographynophoto}
{MIKE BROOKES} is an Emeritus Reader in Signal Processing in the Department of Electrical and Electronic Engineering at Imperial College London. After graduating in Mathematics from Cambridge University in 1972, he worked at the Massachusetts Institute of Technology before returning to the UK and joining Imperial College in 1977.
\end{IEEEbiographynophoto}
\vskip -2.5\baselineskip plus -1fil
\begin{IEEEbiographynophoto}
{JAN ØSTERGAARD} received the M.Sc.E.E. degree from Aalborg University, Aalborg, Denmark, in 1999 and the Ph.D. degree (cum laude) from the Delft University of Technology, Delft, The Netherlands, in 2007. He was an R\&D Engineer with ETI Inc., Chantilly, VA, USA. Between September 2007 and June 2008, he was a Postdoctoral Researcher with The University of Newcastle, Callaghan, NSW, Australia. He is currently a Professor in information theory and signal processing, the Head of the Section on AI and Sound, and the Head of the Centre for Acoustic Signal Processing Research with Aalborg University.
\end{IEEEbiographynophoto}
\vskip -2.5\baselineskip plus -1fil
\begin{IEEEbiographynophoto}
{PATRICK A. NAYLOR} received the B.Eng. degree in electronic and electrical engineering from the University of Sheffield, Sheffield, U.K., and the Ph.D. degree from Imperial College London, London, U.K. He is currently a Professor of speech and acoustic signal processing with Imperial College London. His research interests include speech, audio and acoustic signal processing. His current research addresses microphone array signal processing, speaker diarization, and multichannel speech enhancement for application to binaural hearing aids and robot audition. He has also worked on speech dereverberation including blind multichannel system identification and equalization, acoustic echo control, non-intrusive speech quality estimation, and speech production modeling with a focus on the analysis of the voice source signal. In addition to his academic research, he enjoys several collaborative links with industry.
\end{IEEEbiographynophoto}
\vskip -2.5\baselineskip plus -1fil
\begin{IEEEbiographynophoto}
{JESPER JENSEN} received the M.Sc. degree in electrical engineering and the Ph.D. degree in signal processing from Aalborg University, Aalborg, Denmark, in 1996 and 2000, respectively. From 1996 to 2000, he was with Center for Person Kommunikation, Aalborg University, as a Ph.D. student and an Assistant Research Professor. From 2000 to 2007, he was a Postdoctoral Researcher and an Assistant Professor with the Delft University of Technology, Delft, The Netherlands, and an External Associate Professor with Aalborg University. He is currently a Fellow with Demant A/S, Smørum, Denmark, where his main responsibility is scouting and development of new signal processing concepts for hearing aid applications. He is a Professor with the Section for Signal and Information Processing, Department of Electronic Systems, Aalborg University. He is also a Co-Founder of Centre for Acoustic Signal Processing Research, Aalborg University. His main research interests include acoustic signal processing, including signal retrieval from noisy observations, coding, speech, and audio modification and synthesis, intelligibility enhancement of speech signals, signal processing for hearing aid applications, and perceptual aspects of signal processing.
\end{IEEEbiographynophoto}

\vfill

\end{document}